\documentclass[letterpaper,times]{IONconf}
\usepackage[round,authoryear]{natbib}

\usepackage{amsmath}
\usepackage{amssymb}
\usepackage{bm}

\usepackage{graphicx}
\usepackage[justification=centering]{caption}
 \usepackage{subcaption}
\usepackage{algorithmic}

\usepackage{algorithm}

\usepackage{array}

\usepackage{url}

\usepackage[hidelinks]{hyperref}

\title{Coordination of Ground-to-Space Reference Networks for High-Precision GNSS}

\author{
Xue Xian Zheng$^{1}$,
Xing Liu$^{2,3}$,
José A. López-Salcedo$^{2,3}$,
Gonzalo Seco-Granados$^{2,3}$,
and Tareq Y. Al-Naffouri$^{1}$\\[1ex]
$^{1}$ King Abdullah University of Science and Technology (KAUST), Thuwal, Saudi Arabia\\
$^{2}$ Universitat Autònoma de Barcelona (UAB), Barcelona, Spain\\
$^{3}$ Institut d'Estudis Espacials de Catalunya (IEEC), Barcelona, Spain
}

\begin{document}

\maketitle

\section*{biography}


\biography{Xue Xian Zheng}{is a doctoral student at King Abdullah University of Science and Technolog (KAUST), Thuwal, Saudi Arabia. He is currently specializing in signal processing and optimization methods over graphs, applying these frameworks to advance large-scale satellite navigation.}

\biography{Xing Liu}{is a research scientist with the Institute of Space Studies of Catalonia (IEEC), Barcelona, Spain. He is also with Universitat Autònoma de Barcelona (UAB), Barcelona, Spain. His research interests include GNSS and LEO satellites for positioning, navigation, and timing.}

\biography{José A. López-Salcedo}{is a professor with Universitat Autònoma de Barcelona (UAB), and also with the Institute of Space Studies of Catalonia (IEEC), Barcelona, Spain. His research interests include signal processing for communications and navigation, and the convergence among GNSS, 5G/6G, and LEO satellites for positioning, navigation, and timing.}

\biography{Gonzalo Seco-Granados}{is a professor with Universitat Autònoma de Barcelona (UAB), and also with the Institute of Space Studies of Catalonia (IEEC), Barcelona, Spain. His research interests include signal processing and signal/receiver design for GNSS, low Earth orbit positioning, navigation, and timing, beyond 5G-integrated communications, localization, and sensing.}

\biography{Tareq Y. Al-Naffouri}{is a professor with King Abdullah University of Science and Technolog (KAUST), Thuwal, Saudi Arabia. His research interests include statistical inference and learning, with applications in wireless communications, positioning, satellite navigation, smart cities, and smart health.}

\section*{Abstract}

High-precision Global Navigation Satellite System (GNSS) services rely on accurate orbit, clock, atmospheric, and hardware-bias corrections generated from reference observations. These products are traditionally derived from terrestrial reference networks, whose performance strongly depends on the density and geographic distribution of ground stations. Consequently, sparse or regionally concentrated networks can suffer from tracking gaps and limited global observability, reducing their ability to support globally consistent high-precision products. Low Earth Orbit (LEO) constellations equipped with onboard GNSS receivers and inter-satellite links (ISLs) can serve as a network of spaceborne reference stations, offering a promising way to extend terrestrial reference networks into space, thereby improving observability for ground networks and enabling future direct correction broadcast. However, most existing network-based GNSS correction-generation workflows assume that observations can be centrally collected and processed. This assumption fails in practical ground--to--space architectures, where dynamic satellite geometry and system constraints render communication links intermittent, asymmetric, capacity-limited, and lossy. 

To address these limitations, this paper proposes a decentralized processing architecture that coordinates the ground-to-space GNSS reference network. By modeling ground stations and LEO satellites as interacting subnetworks over a dynamic graph, our approach allows frequent intra-tier communication while restricting cross-tier exchanges to compact estimation summaries transmitted opportunistically under probabilistic link availability. This architecture supports scalable global correction generation, robust LEO orbit/clock estimation, and resilient Precise Point Positioning (PPP) as well as PPP with Real-Time Kinematic (PPP-RTK) services--entirely bypassing the need for an ultra-dense global ground infrastructure. Numerical results show that the proposed method maintains stable estimation and positioning performance under intermittent and lossy communication, improves robustness over ground-only and centrally dependent baselines, and substantially reduces cross-tier communication requirements.

\section{Introduction}

High-precision positioning, navigation, and timing (PNT) in the Global Navigation Satellite System (GNSS) increasingly depends on correction products rather than on broadcast ephemerides alone. Precise Point Positioning (PPP) estimates a user's position from undifferenced carrier-phase and code observations by applying precise satellite orbit and clock products, together with models for atmospheric delay, antenna phase centers, relativity, tides, and hardware biases~\citep{Zumberge1997,Kouba2001}. PPP with real-time kinematic augmentation (PPP-RTK) further reduces convergence time by broadcasting state-space corrections for satellite orbits, clocks, phase/code biases, and atmospheric delays~\citep{Wubbena2005}. These correction services now become a foundational capability not only for geodesy and timing transfer, but also for robotics, intelligent transportation, precision agriculture, Earth science, and safety-critical infrastructure.

The backbone of such services is the terrestrial reference network. The International GNSS Service (IGS) and its Analysis Centers combine globally distributed observations to generate ultra-rapid, rapid, and final products for satellite orbits, clocks, and Earth rotation parameters~\citep{Dow2009,Johnston2017}. When the ground network is dense, well calibrated, and globally distributed, it provides the geometric diversity required to estimate GNSS satellite states and clocks with high accuracy. However, this architecture has a structural limitation: it observes the GNSS space segment only from the Earth's surface. Station geometry is sparse over oceans, polar regions, deserts, conflict regions, and areas where operators cannot deploy or control infrastructure. This matters because the quality of GNSS orbit and clock products is strongly coupled to the number and distribution of tracking stations~\citep{Huang2020,Li2024RegionalLEO}. For regional service providers, commercial constellations, and navigation systems whose monitoring infrastructure is concentrated in one geographic region, the conventional path to global precision---building and operating a dense global ground network---is expensive, politically constrained, and operationally fragile.

Low Earth Orbit (LEO) constellations change this design space. A LEO satellite equipped with a geodetic-grade GNSS receiver can act as a moving spaceborne reference station. Unlike a fixed ground receiver, it rapidly traverses different viewing geometries, observes GNSS satellites from above the atmosphere, and can help bridge tracking gaps left by sparse terrestrial stations. This idea has appeared in combined ground-to-space Precise
Orbit Determination (POD) studies, sometimes called one-step or integrated POD, in which ground and onboard GNSS observations are processed together~\citep{Hugentobler2005,Geng2008,Huang2020}. Recent results make the opportunity concrete. Li \emph{et al.} integrated six regional IGS stations around China with thirteen LEO onboard receivers and showed that regional-only GPS orbit determination, which produced meter-level errors, could be improved to centimeter-level integrated orbit products; the reported average GPS orbit RMS values were 2.27~cm, 3.45~cm, and 3.08~cm in the radial, along-track, and cross-track directions, respectively, with GPS clock accuracy better than 0.15~ns~\citep{Li2024RegionalLEO}. The same study obtained centimeter-level LEO orbit accuracy and demonstrated comparable simulated kinematic PPP performance using the integrated products~\citep{Li2024RegionalLEO}.

This result suggests a powerful architectural shift: a regional ground network can be made globally observable by adding a mobile orbital layer. Yet it also exposes a remaining systems problem. When no receiver observes a GNSS satellite, its clock offset cannot be continuously estimated; clock discontinuities can force ambiguity resets and degrade user positioning~\citep{Li2024RegionalLEO}. Adding more LEO satellites can reduce such gaps, but a large LEO layer does not automatically produce an operational service. The system must also coordinate many observations, clocks, biases, and orbits across a heterogeneous ground-to-space communication fabric.


Furthermore, LEO platforms may also carry inter-satellite links (ISLs), allowing satellites to exchange data, synchronize clocks, and obtain precise relative range constraints. For LEO constellation POD, three observation resources are becoming available: onboard GNSS observations, ISL measurements, and LEO-to-ground navigation or tracking signals~\citep{Li2019ISL,He2022LEOISL,Li2025IntegratedLEOPOD}. A recent 120-satellite simulation showed that adding ground observations to a traditional onboard-GNSS solution improved LEO orbit accuracy by 9.5\%, adding ISLs improved it by 63.1\%, and combining onboard GNSS, ISL, and ground observations improved accuracy by 64.9\% relative to the GNSS-only case~\citep{Li2025IntegratedLEOPOD}. These findings are important because they show that ISLs are not merely communication links: they are also strong geometric constraints that directly improve orbit and clock estimation. They also show that multiple observation types can compensate for limited GNSS availability, regional ground coverage, or temporary sensor outages~\citep{He2022LEOISL,Li2025IntegratedLEOPOD}.

However, most integrated POD formulations still inherit a centralized processing assumption: observations are collected at a fusion center, normal equations are assembled centrally, and global states are solved in batch or near-real time. This assumption is natural for offline geodetic analysis, but it is increasingly mismatched to an operational integrated ground--to--space network. A future network may contain hundreds or thousands of LEO satellites, many regional ground stations, multiple GNSS constellations, ISLs, onboard clocks, atmospheric states, bias states, and user-facing correction broadcast channels. Streaming all raw measurements to a central processor would create high communication overhead, link-schedule contention, stale state updates, and a single point of architectural fragility.

The root cause is communication asymmetry. Ground stations can often communicate with each other through reliable terrestrial backhaul. LEO satellites within the same constellation may communicate frequently through planned ISLs. By contrast, ground-to-space links are limited by orbital visibility, antenna geometry, scheduling, weather, interference, energy, and regulatory constraints. The orbital edge computing literature has shown that bent-pipe space architectures, in which satellites downlink raw data to the ground for processing, do not scale with constellation population~\citep{Denby2020OEC}. Denby and Lucia report an example nanosatellite mission with 88\% packet loss and show that link availability, bitrate, ground-station placement, and contention dominate system capability~\citep{Denby2020OEC}. In their motivating example, an optimistic 1000-satellite bent-pipe constellation requires 112 ideally positioned ground stations, whereas onboard processing that downlinks compact information can reduce this requirement to six~\citep{Denby2020OEC}. Although that work studies Earth observation rather than navigation, the lesson applies directly: a large ground-to-space reference network must process information near where it is observed and communicate compact summaries rather than raw data whenever possible.

This paper brings this systems lesson into high-precision satellite navigation. 
We argue that the next step beyond integrated POD is \emph{decentralized ground-to-space reference networking}, in which terrestrial reference stations, LEO satellites acting as spaceborne reference stations, and GNSS satellites are treated not as passive data sources for a central processor, but as interacting components of a multi-layer dynamic graph. 
In this graph, ground stations form a terrestrial subnetwork, LEO satellites form an orbital subnetwork, and GNSS satellites provide the shared space-segment states to be estimated. 
The communication structure is inherently asymmetric. 
Within each tier, information can be exchanged relatively frequently: ground stations can share products through terrestrial backhaul, while LEO satellites can propagate state information through ISLs. 
Across tiers, however, ground-to-space communication is intermittent and probabilistic; a cross-tier link exists only during visibility windows and succeeds only with a link-dependent probability. 
The estimator should therefore be designed around this natural asymmetry: high-rate intra-tier consistency maintains local stability within the ground and orbital subnetworks, while low-rate cross-tier reconciliation aligns their shared estimates of GNSS orbits, clocks, biases, and other global correction states. 

Our proposed architecture operationalizes this principle by performing local sequential estimation and consensus within each tier. The terrestrial tier maintains station coordinates, receiver clocks, atmospheric states (tropospheric and ionospheric), and ground-derived GNSS corrections. In parallel, the orbital tier manages LEO orbit and clock states, ISL constraints, spaceborne biases, and onboard-derived GNSS corrections. Crucially, during ground-space contact windows, nodes do not bottleneck the network with raw observation data. Instead, they exchange compact inference objects, such as state increments and consensus variables. In this sense, the proposed system acts as a navigational analogue to orbital edge computing. It pushes computation close to the source, communicates only what is strictly necessary for global consistency, and schedules cross-tier data exchange around physical visibility and link constraints.

This framing also highlights the central algorithmic challenge. Recent studies have scaled decentralized GNSS architectures to big-data settings and large constellations, but they generally consider either ground-only or space-only networks \citep{hou2023decentralized,zheng2025decentralized,liu2025leo}. A unified framework for coordinating ground and space segments remains lacking because the two tiers have fundamentally different communication capabilities and link availability. Semi-decentralized federated learning (SD-FL) partly addresses this asymmetry by partitioning nodes into subnetworks, allowing each subnetwork to perform multiple rounds of local information mixing on its own timescale, and sampling representatives for global aggregation \citep{lin2021semi,yemini2022semi}. Although this design balances each subset and reduces the overall communication overhead, it still requires a persistent central aggregator, which is ill-suited to ground-to-space networks because no single node can be assumed to remain continuously reachable. We therefore model the system as a fully decentralized network of interacting subnetworks over a dynamic graph, eliminating the central server. The proposed architecture preserves frequent intra-tier communication as SD-FL does while restricting cross-tier exchanges to compact estimation summaries transmitted opportunistically over probabilistic links, thereby supporting next-generation PPP/PPP-RTK services without requiring an ultra-dense global ground infrastructure.

This paper makes the following contributions.

\textbf{First,} we develop an integrated ground-to-space observation model with an estimable parameterization and formulate GNSS correction-product generation as a cooperative decentralized estimation problem over interacting subnetworks. The formulation distributes observations and state variables across ground and space nodes, scales with both network and constellation size, and removes the centralized data-aggregation assumption commonly adopted in POD and correction-generation systems.

\textbf{Second,} we identify communication asymmetry and intermittent cross-tier connectivity as first-order constraints in operational ground-to-space GNSS integration. To address them, we propose a robust fully decentralized algorithm that supports frequent communication within the ground and space tiers, less frequent communication between tiers, and probabilistic link failures as part of normal system operation. Cross-tier messages contain compact estimation summaries rather than raw observation streams, substantially reducing bandwidth and connectivity requirements.

\textbf{Third,} we establish a simulation framework combining a regional or sparse ground reference network with a LEO constellation to evaluate the proposed method under realistic observation geometry, communication schedules, and stochastic link availability. The experiments assess estimation accuracy, convergence, communication cost, and robustness to link interruptions, and compare the proposed architecture with centralized baselines. The results demonstrate that the proposed framework improves scalability and robustness without sacrificing accuracy.

\section{Methodology}
This section develops the proposed decentralized ground-to-space reference-network framework. First, we introduce the formulation for the integrated ground-to-space GNSS observation model. Second, we detail the communication graph model that governs data exchange across this network. Finally, we present the proposed decentralized computing algorithm.

\subsection{Integrated GNSS Observation Model}
\label{subsec:integrated_gnss_obs}

We consider an integrated observation model in which GNSS satellites are jointly observed by ground reference receivers and onboard GNSS receivers on LEO satellites. The coordinates of the ground reference receivers are assumed to be known, thereby defining the terrestrial reference frame. The LEO and GNSS states are estimated relative to this datum.
Let $\mathcal{V}_{\mathrm{G}}$ denote the set of ground receivers, and $\mathcal{V}_{\mathrm{L}}$ denote the set of LEO onboard receivers. 
At epoch $i$, for a ground receiver $g\in\mathcal{V}_{\mathrm{G}}$, a LEO receiver $\ell\in\mathcal{V}_{\mathrm{L}}$, a visible GNSS satellite $s$, and frequency $j$, the linearized observed-minus-computed undifferenced carrier-phase and pseudo-range observations are modeled as
\begin{equation}
\begin{aligned}
\mathrm{E}\{\Delta \phi_{g,j}^{s}(i)\}
=~& 
-\mathbf{u}_{g}^{s}(i)^{\mathrm{T}}
\Delta \mathbf{p}^{s}(i)
+\mathrm{d}t_{g}(i)-\mathrm{d}t^{s}(i)                                      + \lambda_j \left(\delta_{g,j}(i)-\delta_{j}^{s}(i)+  N_{g,j}^{s} \right) +m_{g}^{s}(i)\tau_{g}(i)
-\mu_j I_{g}^{s}(i), 
\\[0.5em]
\mathrm{E}\{\Delta \rho_{g,j}^{s}(i)\}
=~& 
-\mathbf{u}_{g}^{s}(i)^{\mathrm{T}}
\Delta \mathbf{p}^{s}(i)
+\mathrm{d}t_{g}(i)-\mathrm{d}t^{s}(i)                            
+b_{g,j}(i)-b_{j}^{s}(i)
+m_{g}^{s}(i)\tau_{g}(i)
+\mu_j I_{g}^{s}(i),
\\[0.5em]
\mathrm{E}\{\Delta \phi_{\ell,j}^{s}(i)\}
=~& 
\mathbf{u}_{\ell}^{s}(i)^{\mathrm{T}}
\!\left(\Delta \mathbf{p}_{\ell}(i)-\Delta \mathbf{p}^{s}(i)\right)
+\mathrm{d}t_{\ell}(i)-\mathrm{d}t^{s}(i) + \lambda_j \left(\delta_{\ell,j}(i)-\delta_{j}^{s}(i)+  N_{\ell,j}^{s} \right)
-\mu_j I_{\ell}^{s}(i),
\\[0.5em]
\mathrm{E}\{\Delta \rho_{\ell,j}^{s}(i)\}
=~& 
\mathbf{u}_{\ell}^{s}(i)^{\mathrm{T}}
\!\left(\Delta \mathbf{p}_{\ell}(i)-\Delta \mathbf{p}^{s}(i)\right)
+\mathrm{d}t_{\ell}(i)-\mathrm{d}t^{s}(i)+b_{\ell,j}(i)-b_{j}^{s}(i)
+\mu_j I_{\ell}^{s}(i),
\end{aligned}
\label{eq:integrated_gnss_obs}
\end{equation}
where $\mathrm{E}\{\cdot\}$ denotes expectation, and 
\begin{itemize}
    \item $\Delta \phi_{r,j}^{s}(i)$ and $\Delta \rho_{r,j}^{s}(i)$ denote the observed-minus-computed undifferenced carrier-phase and pseudo-range observations, respectively, for receiver $r \in \{g,\ell\}$, where $r$ denotes either a ground receiver or a LEO receiver.


    \item $\mathbf{u}_{r}^{s}(i)$ is the line-of-sight unit vector from GNSS satellite $s$ to receiver $r$. 

    \item $\Delta\mathbf{p}_{\ell}(i)$ and $\Delta\mathbf{p}^{s}(i)$ denote the position increments of the LEO receiver and the GNSS satellite, respectively.

    \item $\mathrm{d}t_{r}(i)$ and $\mathrm{d}t^{s}(i)$ denote the receiver and GNSS satellite clock offsets expressed in length units. 

    \item $\delta_{r,j}(i)$ and $\delta_{j}^{s}(i)$ denote the receiver-side and satellite-side carrier-phase hardware biases on frequency $j$ expressed in cycles. The terms $b_{r,j}(i)$ and $b_{j}^{s}(i)$ denote the corresponding pseudo-range hardware biases expressed in length units.

    \item $N_{r,j}^{s}$ denotes the undifferenced carrier-phase ambiguity between receiver $r$ and GNSS satellite $s$ on frequency $j$. 

    \item $\tau_{g}(i)$ is the zenith tropospheric delay at ground receiver $g$, and $m_{g}^{s}(i)$ is the corresponding tropospheric mapping function. This term is absent for LEO receivers because they operate above the troposphere.

    \item $I_{r}^{s}(i)$ denotes the slant ionospheric delay on the reference frequency. Its frequency-dependent scaling factor is $\mu_j=\frac{\lambda_j^2}{\lambda_1^2}$, where $\lambda_j$ is the wavelength of frequency $j$ and $\lambda_1$ is the wavelength of the reference frequency. 
\end{itemize}

It is worth noting that \eqref{eq:integrated_gnss_obs} represents a baseline integrated observation model that includes GNSS observations from both ground and LEO onboard receivers. ISL measurements between LEO satellites can be incorporated into \eqref{eq:integrated_gnss_obs}, providing additional geometric constraints and thereby improving estimation accuracy as suggested by \citep{Li2019ISL,Li2025IntegratedLEOPOD}. Regardless of how these measurements are organized, the parameter space of the integrated model can be partitioned into local and shared parameters. For example, after suppressing the epoch index $i$, let $\mathbf{z}$ collect the shared GNSS space-segment parameters, including GNSS satellite orbit errors, clock bias, and hardware biases by $ \mathbf{z}=\{\Delta\mathbf{p}^{s}, \mathrm{d}t^{s}, \delta_{j}^{s}, b_{j}^{s},\dots \}$.
All remaining parameters are collected into the local receiver-related vector $\mathbf{x}_{r\in \{g,\ell\}}$. Under this partition, \eqref{eq:integrated_gnss_obs} can be expressed as
\begin{equation}
\mathbf{y}_r = \mathbf{A}_r \mathbf{x}_r + \mathbf{B}_r \mathbf{z} + \mathbf{n}_r,
\label{eq:obs_r}
\end{equation}
where $\mathbf{y}_r$ denotes the observed-minus-computed observation vector  collected from the L.H.S of \eqref{eq:integrated_gnss_obs}, $\mathbf{A}_r$ and $\mathbf{B}_r$ are the design matrices, and $\mathbf{n}_r$ represents measurement noise. The noise is assumed to have zero mean, $\mathbb{E}(\mathbf{n}_r)=\mathbf{0}$, and covariance $\mathbf{Q}_r$, with $\mathbb{C}(\mathbf{n}_r,\mathbf{n}_{r'})=\delta_{rr'}\,\mathbf{Q}_r$, where $\delta_{rr'}$ denotes the Kronecker delta.

The ground-to-space GNSS network aims to cooperatively obtain each $\mathbf{x}_{r}$ and shared $\mathbf{z}$  through local computation at each node and compact information exchange with neighboring nodes. This avoids uploading all raw observations to a central processor and solving a single large-scale estimation problem. Accordingly, the overall estimation problem can be formulated as
\begin{equation}
\begin{aligned}
& \underset{\{\mathbf{x}_r\}_{r\in \mathcal{V}_{\mathrm{G}}\cup\mathcal{V}_{\mathrm{L}} },\ \mathbf{z}}{\text{minimize}} \quad f(\{\mathbf{x}_r\}_{r\in \mathcal{V}_{\mathrm{G}}\cup\mathcal{V}_{\mathrm{L}}},\mathbf{z})=\sum_{r\in \mathcal{V}_{\mathrm{G}}\cup\mathcal{V}_{\mathrm{L}}}f_{r}(\mathbf{x}_{r},\mathbf{z}),\\
 &\quad \mathrm{where} \quad \quad f_{r}(\mathbf{x}_{r},\mathbf{z}) = \frac{1}{2}\big\|\,\mathbf{A}_{r}\mathbf{x}_{r}+\mathbf{B}_{r}\mathbf{z}-\mathbf{y}_r\,\big\|_{\mathbf{Q}_{r}^{-1}}^2.
\end{aligned}
\label{eq:obs_dec}
\end{equation}

\textbf{Remark 1} (Identifiability and estimability). \textit{Formulation \eqref{eq:obs_dec} is mathematically equivalent to the centralized weighted least-squares formulation obtained by stacking all receiver observations, provided that local copies of the shared parameter vector $\mathbf{z}$ are constrained to reach consensus. Nevertheless, the estimation problem may be rank deficient. In that case, each local objective $f_{r}(\mathbf{x}_{r},\mathbf{z})$ can be ill posed because only certain linear combinations of the parameters are estimable, while their absolute values remain undetermined. Hence, all nodes must employ a common estimable parameter basis. In the integrated ground-to-space GNSS network, such a basis can be constructed using $\mathcal{S}$-system theory with a ground-oriented reference datum \citep{teunissen2010ppp,odijk2016estimability,hou2023decentralized}. Specifically, selected datum-related parameters in $\mathbf{x}_r$ are fixed for reference ground stations $r=g\in\mathcal{V}_{\mathrm{G}}$, thereby making the corresponding $f_{r}(\mathbf{x}_{r},\mathbf{z})$ well posed. The estimable satellite parameters observed by these reference stations then define a common reference frame. This reference is subsequently transferred to LEO receivers through common-view GNSS observations, enabling consistent estimation of LEO orbit errors, clock parameters, and hardware biases for $r=\ell\in\mathcal{V}_{\mathrm{L}}$. Moreover, LEO receivers may observe additional GNSS satellites that are not directly visible to the reference ground stations. Through these common-view links, the datum information can propagate through the whole ground-to-space network, thereby enabling convergence to a unique estimable solution.}

\subsection{Communication Graph Model}

\begin{figure*}[!t]
\centering
\includegraphics[width=1\textwidth]{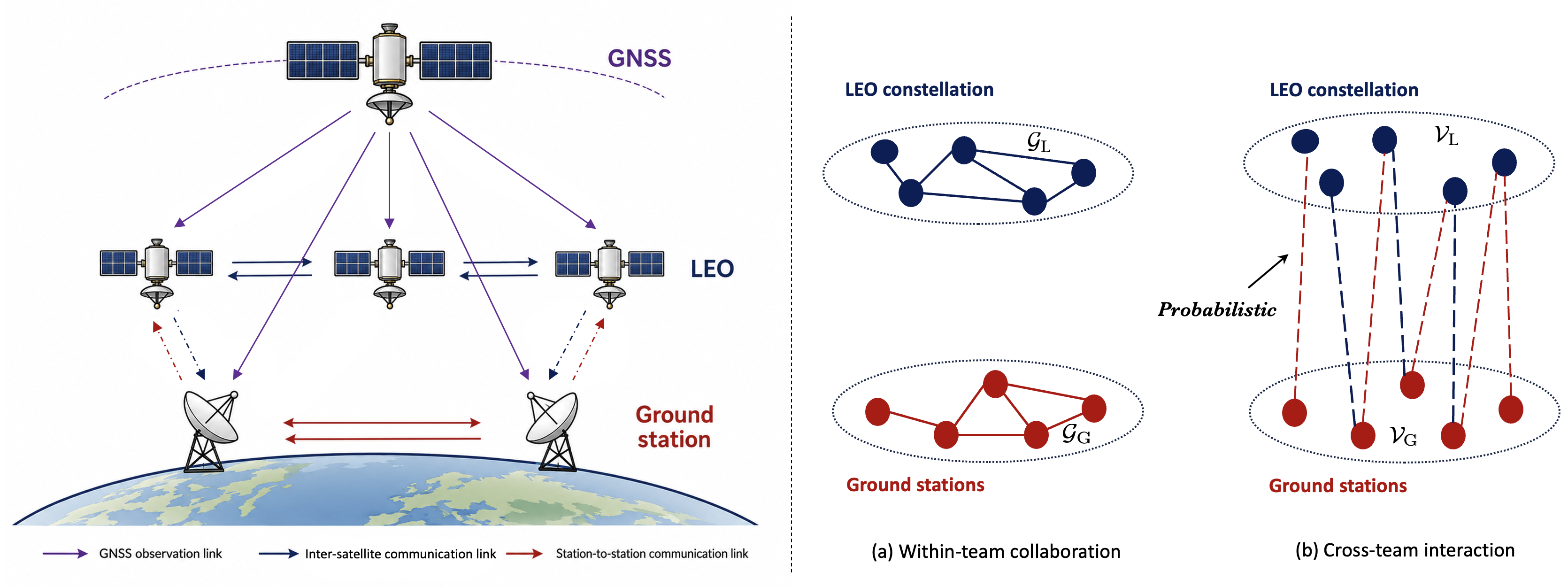}
\caption{{\bf{(Left.)}} Physical communication pattern in the ground-to-space GNSS network. Ground stations and LEO satellites cooperatively process GNSS observations. Ground stations communicate via reliable terrestrial backhaul, while LEO satellites exchange information through planned ISLs. Ground-to-space communication links, however, are opportunistic and subject to probabilistic availability. \textbf{(Right.)} Cooperation strategy for distributed estimation. Nodes within the same group communicate frequently according to their own time scales, whereas cross-tier exchanges are restricted to compact estimation summaries and occur less frequently under probabilistic link availability.}
\label{fig_1}
\end{figure*}
While the careful selection of the estimable parameter basis and the local construction of the design matrices $\mathbf{A}_r$ and $\mathbf{B}_r$ address the rank-deficiency issue, decentralized processing further depends on the communication topology. The communication graph of an integrated ground-to-space GNSS network is fundamentally different from that of a ground-only or space-only network. As it is shown in Fig. \ref{fig_1} left,  the network naturally consists of two communication tiers: the ground tier and the LEO tier. The ground tier is usually supported by terrestrial backhaul links, whereas the LEO tier communicates through planned inter-satellite links. In contrast, ground-to-space communication is limited by visibility, antenna pointing, scheduling, weather, interference, and energy constraints. Therefore, a realistic decentralized algorithm should exploit frequent within-tier communication while tolerating random and possibly failed cross-tier links.

Let $\mathcal{V}=\mathcal{V}_{\mathrm{G}}\cup \mathcal{V}_{\mathrm{L}}$ denote the set of all computing nodes $r \in \{g,\ell\}$, with $n_{\mathrm{G}}=|\mathcal{V}_{\mathrm{G}}|$, $n_{\mathrm{L}}=|\mathcal{V}_{\mathrm{L}}|$, and $n=n_{\mathrm{G}}+n_{\mathrm{L}}$. The ground and LEO intra-tier communication graphs are denoted by
$\mathcal{G}_{\mathrm{G}}=(\mathcal{V}_{\mathrm{G}},\mathcal{E}_{\mathrm{G}}, \mathbf{W}_{\mathrm{G}})$ and $
\mathcal{G}_{\mathrm{L}}=(\mathcal{V}_{\mathrm{L}},\mathcal{E}_{\mathrm{L}},\mathbf{W}_{\mathrm{L}})$, respectively. Here, $\mathcal{E}_{\mathrm{G}}\subseteq\mathcal{V}_{\mathrm{G}}\times\mathcal{V}_{\mathrm{G}}$ and $\mathcal{E}_{\mathrm{L}}\subseteq\mathcal{V}_{\mathrm{L}}\times\mathcal{V}_{\mathrm{L}}$ denote the corresponding edge sets. To orchestrate information flow, the mixing matrices $\mathbf{W}_{\mathrm{G}}=[w_{gp}]\in\mathbb{R}^{n_{\mathrm{G}}\times n_{\mathrm{G}}}$ and $\mathbf{W}_{\mathrm{L}}=[w_{\ell q}]\in\mathbb{R}^{n_{\mathrm{L}}\times n_{\mathrm{L}}}$ respect these network topologies, meaning that $w_{gp}>0$ only if $(g,p)\in\mathcal{E}_{\mathrm{G}}$ or $g=p$; similarly, $w_{\ell q}>0$ only if $(\ell,q)\in\mathcal{E}_{\mathrm{L}}$ or $\ell=q$. We further assume these graphs are connected over their own tiers, and both $\mathbf{W}_{\mathrm{G}}$ and $\mathbf{W}_{\mathrm{L}}$ are doubly stochastic, where
\begin{equation}
\mathbf{W}_{\mathrm{G}}\mathbf{1}=\mathbf{1},\quad
\mathbf{1}^{\mathrm{T}}\mathbf{W}_{\mathrm{G}}=\mathbf{1}^{\mathrm{T}},
\qquad
\mathbf{W}_{\mathrm{L}}\mathbf{1}=\mathbf{1},\quad
\mathbf{1}^{\mathrm{T}}\mathbf{W}_{\mathrm{L}}=\mathbf{1}^{\mathrm{T}}.
\label{eq:doubly}
\end{equation}
While prior works have studied time-varying $\mathbf{W}_{\mathrm{G}}$ and $\mathbf{W}_{\mathrm{L}}$ in ground-only \citep{zheng2025decentralized} and space-only \citep{liu2025leo} systems—accounting for schedule optimization and orbital dynamics, respectively—we simplify our model by treating these intra-tier mixing matrices as static. We apply them multiple times during the intra-tier communication phase. During this within-tier processing phase, the overall communication graph is characterized by the block-diagonal matrix:
\begin{equation}
\mathbf{W}_{\mathrm{in}}
=
\begin{bmatrix}
\mathbf{W}_{\mathrm{G}} & \mathbf{0}\\
\mathbf{0} & \mathbf{W}_{\mathrm{L}}
\end{bmatrix} \in \mathbb{R}^{n \times n}, \quad \mathbf{W}_{\mathrm{in}}\mathbf{1}=\mathbf{1},\quad
\mathbf{1}^{\mathrm{T}}\mathbf{W}_{\mathrm{in}}=\mathbf{1}^{\mathrm{T}}.
\label{eq:Win}
\end{equation}
This block-diagonal structure ensures that ground nodes exchange information solely with other ground nodes, and LEO nodes communicate exclusively with other LEO nodes.

At the outer communication round $t$, the cross-tier graph is modeled as a random bipartite graph
\begin{equation}
\mathcal{G}_{\mathrm{b}}^{t}
=
\big(\mathcal{V}_{\mathrm{G}},\mathcal{V}_{\mathrm{L}},
\mathcal{E}_{\mathrm{b}}^{t}\big), \quad \textnormal{where} \quad \mathcal{E}_{\mathrm{b}}^{t}\subseteq \mathcal{V}_{\mathrm{G}}\times \mathcal{V}_{\mathrm{L}}.
\label{eq:bipati}
\end{equation}
A candidate cross-tier edge $(g,\ell)$ can be scheduled only when the corresponding ground station $g$ and LEO node $\ell$ are mutually visible and permitted by the communication schedule. Let $\chi_{g\ell}^{t}\in\{0,1\}$ denote this scheduling-and-visibility indicator, and it forms cross-tier adjacency matrix $\mathbf{X}^{t}\in\{0,1\}^{n_{\mathrm{G}}\times n_{\mathrm{L}}}$. When $\chi_{g\ell}^{t}=1$, the LEO node $\ell$ is visible to the ground station $g$ and transmission is scheduled. However, the transmission may still fail with a certain probability due to adverse channel conditions, such as rain, snow, or other weather-related impairments. Therefore, we model the actual link activation by
\begin{equation}
a_{g\ell}^{t}=\chi_{g\ell}^{t}\xi_{g\ell}^{t},
\qquad
\mathbb{P}\{\xi_{g\ell}^{t}=1\}=p_{g\ell},
\label{eq:link_failure}
\end{equation}
where $p_{g\ell}\in[0,1]$ is the success probability of the ground-to-space link. Hence, $(g,\ell)\in\mathcal{E}_{\mathrm{b}}^{t}
 \Longleftrightarrow 
a_{g\ell}^{t}=1.$ By assembling all entries $a_{g\ell}^{t}$, we obtain the actual cross-tier adjacency matrix $\mathcal{A}^{t}\in\{0,1\}^{n_{\mathrm{G}}\times n_{\mathrm{L}}}$. Thus, $\mathcal{A}^{t}$ can be interpreted as a random edge-sampling model obtained by Bernoulli thinning of the scheduled candidate $\mathbf{X}^{t}$ according to the link success probabilities ${p_{g\ell}}$, similar to the random graph models discussed in~\citep{Isufi,zheng2026quantitative}.

Although the link-success probabilities $\{p_{g\ell}\}$ may be unknown, their values are not needed to construct the mixing matrix. Instead, the mixing weights can be determined directly from the realized cross-tier adjacency matrix $\mathcal{A}^{t}$. In particular, we can construct a global adjacency matrix $\mathcal{A}_{\mathrm{b}}^{t}$ and a degree matrix $\mathcal{D}_{\mathrm{b}}^{t}$ by using
\begin{equation}
\mathcal{A}_{\mathrm{b}}^{t}
=
\begin{bmatrix}
\mathbf{0} & \mathcal{A}^{t}\\
{\mathcal{A}^{t}}^\top & \mathbf{0}
\end{bmatrix} \in \{0,1\}^{n \times n},\quad \mathcal{D}_{\mathrm{b}}^{t}=\mathrm{diag}(\mathcal{A}_{\mathrm{b}}^{t}\mathbf{1})=\mathrm{diag}(\{d_{g}^{t}\}_{g\in\mathcal{V}_{\mathrm{G}}},\{d_{\ell}^{t}\}_{\ell\in\mathcal{V}_{\mathrm{L}}}),
\label{eq:metropolis_offdiag}
\end{equation}
where $d_g^{t}$ and $d_\ell^{t}$ are the degree of the ground and LEO node in the active bipartite graph. Then a doubly stochastic symmetric matrix $\mathbf{W}_{\mathrm{b}}^{t}\in\mathbb{R}^{n\times n}$ is constructed according to the Metropolis rule:
\begin{equation}
\quad
[\mathbf{W}_{\mathrm{b}}^{t}]_{g(\ell+n_\mathrm{G})}
=
\frac{1}{1+\max\{d_g^{t},d_\ell^{t}\}}, \quad [\mathbf{W}_{\mathrm{b}}^{t}]_{gg}
=
1-\sum_{l}[\mathbf{W}_{\mathrm{b}}^{t}]_{g(\ell+n_\mathrm{G})}, \quad [\mathbf{W}_{\mathrm{b}}^{t}]_{(\ell+n_\mathrm{G}) (\ell+n_\mathrm{G})}
=
1-\sum_{g}[\mathbf{W}_{\mathrm{b}}^{t}]_{g(\ell+n_\mathrm{G})}
\label{eq:metropolis_diag}
\end{equation}
where $\ell$ specifies the neighbors of $g$, and all off-diagonal entries associated with inactive or nonexistent links are zero. Then, $\mathbf{W}_{\mathrm{b}}^{t}\mathbf{1}=\mathbf{1}$ and $
\mathbf{1}^{\mathrm{T}}\mathbf{W}_{\mathrm{b}}^{t}
=
\mathbf{1}^{\mathrm{T}}$ hold. If all cross-tier links fail in round $t$, then $\mathcal{A}_{\mathrm{b}}^{t}=\mathbf{0}$, and the construction reduces to $\mathbf{W}_{\mathrm{b}}^{t}=\mathbf{I}$.

Consequently, a failed cross-tier communication round does not introduce an erroneous mixing update; each node simply retains its current message during that round. Moreover, as indicated by \eqref{eq:metropolis_diag}, the weights can be computed locally from the realized active links using only their endpoint degrees $d_g^{t}$ and $d_\ell^{t}$. The resulting mixing matrix therefore adapts automatically to each realization of the random cross-tier graph, making this construction well suited to the integrated GNSS networks.

\textbf{Remark 2} (Dynamics of the union graph). \textit{This model captures the two-time-scale cooperation pattern illustrated in the right panel of Fig. \ref{fig_1}. Nodes can perform multiple planned updates within their own tiers using $\mathbf{W}_{\mathrm{G}}$ and $\mathbf{W}_{\mathrm{L}}$, while only a single, compact cross-tier update is attempted through the random bipartite matrix $\mathbf{W}_{\mathrm{b}}^{t}$ at the end of each outer round. Note that at either of these two time scales, neither graph exhibits global connectivity on its own. However, through their interaction, the system dynamically transfers information throughout the entire network, which is crucial for the algorithm development.}

\subsection{Decentralized Computing Algorithm}
\label{subsec:decentralized_algorithm}

\begin{algorithm}[!t]
\caption{Nested decentralized gradient tracking for integrated ground-to-space GNSS estimation}
\label{alg:nested_gt}
\begin{algorithmic}[1]
\STATE \textbf{Input:} stepsize $\mu_{g}$, $\mu_{\ell}$, and $\mu_{r}$ , intra-tier mixing matrices $\mathbf{W}_{\mathrm{G}}$ and $\mathbf{W}_{\mathrm{L}}$, inner iteration numbers $K$.
\STATE Each node $r$ initializes $\mathbf{z}_r^{0}$ and sets $\mathbf{g}^{0}_r = \nabla f_r(\mathbf{z}_r^{0})$.
\FOR{$t=0,1,2,\ldots$}
    \STATE Set $\mathbf{z}_r^{t,0}=\mathbf{z}_r^{t}$ and $\mathbf{g}_r^{t,0}=\mathbf{g}_r^{t}$ for all $r\in\mathcal{V}$.
        \FOR{$k=0,\ldots,K-1$}
            \STATE \colorbox{pink!40}{Ground tier $g\in\mathcal{V}_\mathrm{G}$ updates its $\mathbf{z}_g^{t,k}$ and $\mathbf{g}_g^{t,k}$ by \eqref{eq:diffusionGT} with planed mixing $\mathbf{W}_{\mathrm{G}}$}.
            \STATE \colorbox{cyan!20}{LEO tier $\ell\in\mathcal{V}_\mathrm{L}$ updates its $\mathbf{z}_\ell^{t,k}$ and $\mathbf{g}_\ell^{t,k}$ by \eqref{eq:diffusionGT2} with planed mixing $\mathbf{W}_{\mathrm{L}}$}.
        \ENDFOR
    \STATE Collect $\mathbf{z}_r^{t,K}\in \{\mathbf{z}_g^{t,K},\mathbf{z}_\ell^{t,K}\}$ and $\mathbf{g}_r^{t,K}\in \{\mathbf{g}_g^{t,K},\mathbf{g}_\ell^{t,K}\}$ for $r\in\mathcal{V}$.
    \STATE \colorbox{green!20}{By $p_{g\ell}\in[0,1]$, sample the bipartite graph $\mathcal{G}_{\mathrm{b}}^{t}$ according to \eqref{eq:link_failure} and construct $\mathbf{W}_{\mathrm{b}}^{t}$ by \eqref{eq:metropolis_offdiag} and \eqref{eq:metropolis_diag} }.
    \STATE \colorbox{green!20}{Cross-tier $r\in\mathcal{V}$ updates its $\mathbf{z}_r^{t,K}$ and $\mathbf{g}_r^{t,K}$ by \eqref{eq:diffusionGT3} with dynamic and opportunistic mixing $\mathbf{W}_{\mathrm{b}}^{t}$}.
\ENDFOR
\end{algorithmic}
\end{algorithm}

The objective in \eqref{eq:obs_dec} contains both local receiver-specific parameters $\mathbf{x}_{r \in \{g,\ell\}}$ and shared GNSS-related parameters $\mathbf{z}$. Since $\mathbf{x}_{r}$ is needed only by receiver $r$, it is unnecessary to exchange $\mathbf{x}_r$ across the network. Therefore, we let each node locally eliminate its own variable, exchanging only compact summaries associated with the shared parameter vector.

Based on the elimination method from \citep{zheng2025decentralized}, for a given $\mathbf{z}$, we define $\mathbf{x}_r
=
\operatorname{argmin}_{\mathbf{x}_r}
f_r(\mathbf{x}_r,\mathbf{z})$ and reduce $f_r(\mathbf{x}_r,\mathbf{z})$ to $f_r(\mathbf{z})$ as follows: 
\begin{equation}
\begin{aligned}
f_{r}(\mathbf{z})= \frac{1}{2}\big\|\,\mathbf{C}_{r}\big(\mathbf{B}_{r}\mathbf{z}-\mathbf{y}_r\big)\big\|_{\mathbf{Q}_{r}^{-1}}^2, \quad \textnormal{where} \quad
\mathbf{C}_{r}\triangleq \mathbf{I}-\mathbf{A}_{r}(\mathbf{A}^\top_{r}\mathbf{Q}_{r}^{-1}\mathbf{A}_{r})^{-1}\mathbf{A}^\top_{r}\mathbf{Q}_{r}^{-1}.
\end{aligned}
\label{eq:local_x_z}
\end{equation}
Its gradient with respect to $\mathbf z$ is $\nabla f_r(\mathbf z)
= \mathbf B_r^\top \mathbf C_r^\top \mathbf Q_r^{-1}\mathbf C_r\big(\mathbf B_r \mathbf z - \mathbf y_r\big)$. Note that \eqref{eq:local_x_z} is performed in the common estimable parameter basis discussed in Remark 1. Once $\mathbf{z}$ is fixed, every $f_{r}(\mathbf{x}_{r},\mathbf{z})$ in \eqref{eq:obs_dec} becomes well-posed, which establishes the validity of \eqref{eq:local_x_z}.

To ensure convergence to the centralized solution, we introduce a nested decentralized gradient tracking (GT) method. Specifically, each tier performs standard GT for $K$ steps at the smaller time scale, followed by a single cross-tier GT update at the larger time scale $t$. Mathematically, this is expressed as follows. In every outer round $t$, the proposed method executes in two stages. First, for inner iterations $k=0,\dots,K-1$, the nodes update in parallel:
\begin{equation}
\begin{aligned}
     \mathbf{z}_{g}^{t,k+1}= \sum _{p} [\mathbf{W}_{\mathrm{G}}]_{pg}(\mathbf{z}_{p}^{t,k}-\mu_{g}\mathbf{g}^{t,k}_{p}),\quad
      \mathbf{g}_{g}^{t,k+1}=\sum _{p}[\mathbf{W}_{\mathrm{G}}]_{pg}\mathbf{g}^{t,k}_{p} + \nabla f_{g}(\mathbf{z}_{g}^{t,k+1})-\nabla f_{g}(\mathbf{z}_{g}^{t,k}), \quad \textnormal{for} \quad g \in \mathcal{V}_{\mathrm{G}} 
\end{aligned}
 \label{eq:diffusionGT}
\end{equation}
\begin{equation}
\begin{aligned}
     \mathbf{z}_{\ell}^{t,k+1}= \sum _{q} [\mathbf{W}_{\mathrm{L}}]_{q\ell}(\mathbf{z}_{q}^{t,k}-\mu_{\ell}\mathbf{g}^{t,k}_{q}),\quad
      \mathbf{g}_{\ell}^{t,k+1}=\sum _{q}[\mathbf{W}_{\mathrm{L}}]_{q\ell}\mathbf{g}^{t,k}_{q} + \nabla f_{\ell}(\mathbf{z}_{\ell}^{t,k+1})-\nabla f_{\ell}(\mathbf{z}_{\ell}^{t,k}), \quad \textnormal{for} \quad \ell \in \mathcal{V}_{\mathrm{L}}. 
\end{aligned}
 \label{eq:diffusionGT2}
\end{equation}
Second, a cross-tier update is performed for all nodes $r \in \mathcal{V}$:
\begin{equation}
\begin{aligned}
     \mathbf{z}_{r}^{t+1,K}= \sum _{m} [\mathbf{W}_{\mathrm{b}}^{t}]_{mr}(\mathbf{z}_{m}^{t,K}-\mu_{r}\mathbf{g}^{t,K}_{m}),\quad
      \mathbf{g}_{r}^{t+1,K}=\sum _{m}[\mathbf{W}_{\mathrm{b}}^{t}]_{mr}\mathbf{g}^{t,K}_{m} + \nabla f_{r}(\mathbf{z}_{r}^{t+1,K})-\nabla f_{r}(\mathbf{z}_{r}^{t,K}), \quad \textnormal{for} \quad r \in \mathcal{V}, 
\end{aligned}
 \label{eq:diffusionGT3}
\end{equation}
where $m$ indexes the neighbors of $r$. Here, $\mu_{g}$, $\mu_{\ell}$, and $\mu_{r}$ denote the step sizes, and $\mathbf{g}_r$ tracks the global average gradient using neighbor-to-neighbor communication, initialized as $\mathbf{g}^{0,0}_r = \nabla f_r(\mathbf{z}_r^{0,0})$. The overall procedure is summarized in Algorithm \ref{alg:nested_gt}.

Together, this decentralized design matches the physical ground-to-space communication structure: each cluster can run multiple full GT iterations locally, while each outer round performs one full GT iteration across a random bipartite ground-to-space graph with probabilistic link failures. No central processor is required, and raw GNSS observations remain local.

\textbf{Remark 3} (Convergence guarantee). \textit{The key property that enables Algorithm \ref{alg:nested_gt} to converge relies on the observation from Remark 2: the union of the graphs remains connected, even though the individual intra-tier matrices $\mathbf{W}_{\mathrm{G}}$, $\mathbf{W}_{\mathrm{L}}$, and the cross-tier matrix $\mathbf{W}_{\mathrm{b}}^{t}$ are not globally connected on their own. Specifically, the overall dynamics described by \eqref{eq:diffusionGT}, \eqref{eq:diffusionGT2}, and \eqref{eq:diffusionGT3} can be viewed as a large-scale GT process over a global graph, with a mixing sequence given by:
\begin{equation}
    \Big\{\underbrace{\mathbf{W}_{\mathrm{in}},\ \dots,\mathbf{W}_{\mathrm{in}}}_{K},\ \mathbf{W}_{\mathrm{b}}^{0},| \dots,| \underbrace{\mathbf{W}_{\mathrm{in}},\ \dots,\mathbf{W}_{\mathrm{in}}}_{K},\ \mathbf{W}_{\mathrm{b}}^{t},| \dots\Big\}
\label{eq:coverge}
\end{equation}
Consequently, for each large time scale $t$, the product of the mixing matrices $\mathbf{W}^{K}_{\mathrm{in}}\mathbf{W}_{\mathrm{b}}^{t}$ remains contractive as long as the link activation probability satisfies $p_{g\ell} > 0$. By standard GT analysis \citep{nedic2017achieving,zheng2025decentralized}, this contractiveness guarantees convergence, meaning that eventually:
\begin{equation}
    \lim_{t\rightarrow\infty}
\mathbb{E}\left[
\|\mathbf{z}_r^{t}-\mathbf{z}^{\star}\|^2
\right]
=
0,
\qquad
\forall r\in\mathcal{V},
\end{equation}
where $\mathbf{z}^{\star}$ is the global minimizer. The primary distinction in our setting lies in the characterization of the convergence rate, which is inherently dictated by the link activation probability $p_{g\ell}$. An increased $p_{g\ell}$ establishes more cross-tier data pathways, which consequently yields a strictly faster rate of convergence. In other words, the network's vulnerability to cross-tier link failures does not compromise the guarantee of convergence; it solely impacts the rate of convergence. We will corroborate this theoretical finding in our numerical experiments.}

\section{Results}
\begin{figure*}[!t]
\centering
\includegraphics[width=1.01\textwidth]{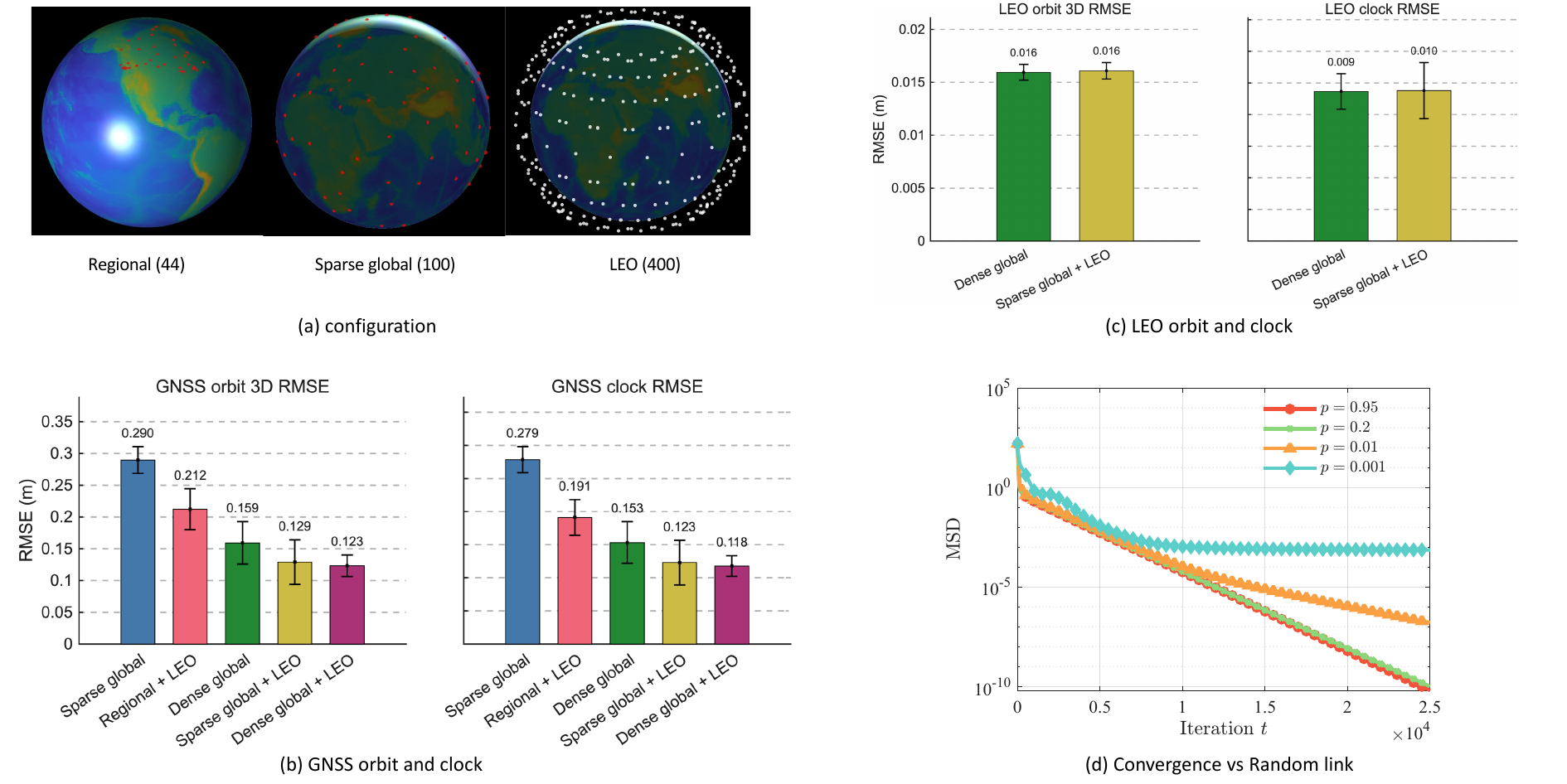}
\caption{Numerical experiment configuration and results. (a) The ground and LEO GNSS networks used in the experiments. (b) The GNSS orbit and clock error estimation results, where \textit{Dense global} corresponds to the full network of 529 IGS stations. (c) The LEO orbit and clock error estimation results. (d) The convergence behavior of the proposed method under different link activation probabilities.}
\label{fig_2}
\end{figure*}

This section presents the numerical results for the proposed architecture. We evaluate the method on a single-epoch ground-to-space GNSS network. This single-epoch approach stands in contrast to the large-batch offline analyses spanning several weeks presented by \cite{Li2024RegionalLEO}. We define the global state $\mathbf{z}=\{\Delta\mathbf{p}^{s}, \mathrm{d}t^{s},\dots\}$ to encompass the orbit error of the GNSS satellite and its clock error, both expressed in meters. In the simulations, only GPS satellites are considered, totaling 30 satellites. Dual-frequency observations at the GPS L1/L2 frequencies are simulated, including undifferenced carrier-phase and code pseudorange measurements. An elevation mask of 0° is applied, and only LOS visible GNSS satellites are used. The carrier-phase standard deviation is set to 0.001 m, and the code standard deviation is 15 times larger (0.015 m). GNSS satellite orbits are assigned an initial uncertainty of 5 m, and clock offsets are assigned an initial uncertainty of 10 ns. Regarding the observation network, two ground-station layouts are evaluated: a localized Canadian IGS network comprising 44 stations, and a sparse, globally distributed IGS network of 100 stations from \citep{zheng2025decentralized}. In both configurations, the initial station serves as the reference datum with a zero clock bias. A Walker-Delta LEO constellation consisting of 400 satellites is considered, where each satellite is equipped with a dual-frequency GNSS receiver. The satellites are deployed in 20 orbital planes with 20 satellites per plane, at an altitude of 1000 km and an inclination of 53°, and their trajectories are propagated accordingly. Together, their distributions are plotted in Fig. \ref{fig_2} (a). All receiver clock offsets are initialized with a standard deviation of 100 ns. Ground reference stations are assumed to have fixed positions, while LEO satellite positions are assigned an initial uncertainty of 5 m. The resulting simulated observation set is used for
the subsequent estimation analysis, where all reported errors are root-mean-square errors (RMSE) in meters.

Regarding the algorithmic settings, we construct the intra-tier communication graphs $\mathcal{G}_{\mathrm{G}}$ and $\mathcal{G}_{\mathrm{L}}$ using a 5-nearest neighbor rule, meaning each node can only communicate with its 5 closest neighbors within its own tier. For the bipartite graph $\mathcal{G}_{\mathrm{b}}^{t}$, we set the cross-tier link probability to $p = p_{g\ell} = 0.95$, which is sufficiently high for simulating orbit and clock product generation. We later detail the tuning of this parameter to evaluate the theoretical convergence guarantees. Furthermore, we set a uniform step size of $\mu_{r}=0.01$ for all nodes $r$, configure the number of inner iterations to $K=10$, and cap the total number of outer rounds at $t \to 2.5 \times 10^4$. All numerical results are given in Fig. \ref{fig_2} (b)-(d).

\subsection{GNSS orbit and clock estimation}

Fig.~\ref{fig_2} (b) evaluates the GNSS satellite orbit and clock estimation accuracy under five representative tracking configurations. Each bar reports the mean RMSE over five independent single-epoch Monte Carlo runs, and the error bars denote the corresponding standard deviation.

The \textit{Sparse global} ground-only network provides the weakest observability, yielding a GNSS orbit 3D RMSE of $0.290~\mathrm{m}$ and a clock RMSE of $0.279~\mathrm{m}$. Increasing the number and geographic diversity of ground stations significantly improves the solution. In the \textit{Dense global} ground-only case, the orbit and clock RMSEs decrease to $0.159~\mathrm{m}$ and $0.153~\mathrm{m}$, respectively, corresponding to reductions of approximately $45.0\%$ for both orbit and clock estimation relative to the sparse global baseline.

LEO augmentation provides an additional and complementary source of GNSS tracking geometry. Even when the ground segment is geographically localized, the \textit{Regional + LEO} configuration achieves orbit and clock RMSEs of $0.212~\mathrm{m}$ and $0.191~\mathrm{m}$, respectively. This represents improvements of $26.8\%$ and $31.4\%$ over the \textit{Sparse global} ground-only case. These results indicate that spaceborne GNSS observations can partially compensate for limited ground coverage by increasing the diversity of line-of-sight directions and strengthening the observability of the GNSS satellite states.

The advantage of LEO augmentation becomes more pronounced when combined with a sparse but globally distributed ground network. The \textit{Sparse global + LEO} configuration reduces the GNSS orbit and clock RMSEs to $0.129~\mathrm{m}$ and $0.123~\mathrm{m}$, respectively. Notably, this performance surpasses that of the \textit{Dense global} ground-only network, despite using far fewer ground stations. Relative to the \textit{Dense global} case, the \textit{Sparse global + LEO} solution improves the orbit and clock RMSEs by approximately $19.0\%$ and $19.8\%$, respectively. This demonstrates that LEO-borne GNSS receivers do more than simply add measurements; they provide geometrically valuable spaceborne tracking links that can reduce the dependence on dense ground infrastructure.

The best GNSS correction performance is achieved by the \textit{Dense global + LEO} configuration, for which the orbit 3D RMSE and clock RMSE reach $0.123~\mathrm{m}$ and $0.118~\mathrm{m}$, respectively. Compared with the \textit{Sparse global} ground-only network, this corresponds to improvements of $57.5\%$ in orbit estimation and $57.7\%$ in clock estimation. Even relative to the \textit{Dense global} ground-only case, the inclusion of LEO observations further reduces the orbit and clock RMSEs by $22.7\%$ and $23.1\%$, respectively. These results show that the proposed ground--LEO architecture provides a substantial enhancement for single-epoch GNSS orbit and clock estimation, with the strongest performance obtained when dense terrestrial tracking and LEO-based spaceborne tracking are jointly exploited.

\subsection{LEO orbit and clock estimation}

Fig.~\ref{fig_2} (c) compares the LEO local orbit and receiver-clock estimation accuracy under two correction-support configurations: \textit{Dense global} and \textit{Sparse global + LEO}. The \textit{Dense global} case uses GNSS orbit and clock corrections estimated from the dense global ground network, whereas the \textit{Sparse global + LEO} case uses the joint solution obtained from the sparse global ground network and the LEO-borne GNSS observations. The results are averaged over five independent single-epoch Monte Carlo runs, and the error bars denote the standard deviation across these runs.

The two configurations produce nearly identical LEO local-state accuracy. With \textit{Dense global} correction support, the LEO orbit 3D RMSE and receiver-clock RMSE are $0.0159~\mathrm{m}$ and $0.00946~\mathrm{m}$, respectively. With \textit{Sparse global + LEO} support, the corresponding RMSEs are $0.0161~\mathrm{m}$ and $0.00952~\mathrm{m}$. This indicates that replacing dense global ground correction support with the proposed sparse-ground--LEO joint architecture does not noticeably degrade the recovered LEO local orbit or receiver-clock states in the present experiment.

The similarity between the two LEO results should be interpreted carefully. The LEO observations do include GNSS orbit and clock errors, as in the ground receiver observation model. However, the LEO receiver does not respond to the separately reported GNSS orbit RMSE and GNSS clock RMSE independently. Instead, each LEO measurement is affected by the combined range-level effect of the GNSS orbit and clock correction errors along the corresponding line of sight. In the current single-epoch least-squares solution, these two GNSS error components are strongly coupled, so their combined effect at the LEO receivers is much smaller than the individual GNSS orbit and clock RMSE values shown in Fig.~\ref{fig_2} (b).

Numerically, although the GNSS orbit and clock RMSEs are at the decimeter level, the residual GNSS correction error projected into the LEO measurements is only at the millimeter-to-centimeter level for both support configurations. As a result, the LEO local-state estimates are dominated primarily by the low assumed code/carrier measurement noise and the LEO viewing geometry. This explains why the LEO orbit and clock bars remain nearly unchanged, even though the GNSS correction products differ more visibly.

Therefore, Fig.~\ref{fig_2} (c) should be viewed as a local single-epoch LEO state-recovery consistency test under the assumed measurement model, rather than as a complete operational LEO precise-orbit-determination assessment. The result shows that the proposed sparse-ground--LEO joint architecture can preserve LEO local orbit and clock recovery accuracy comparable to dense global ground support in this idealized setting. A full LEO POD performance evaluation would require additional modeling of multi-epoch dynamics, process noise, atmospheric residuals, multipath, antenna phase-center effects, attitude uncertainty, and other unmodeled error sources.

\subsection{Convergence under unreliable cross-layer links}
We further evaluate the robustness of the proposed method when the cross-tier ground-to-LEO links operate with a varying link activation probability $p$. In this experiment, we use the \textit{Sparse global + LEO} topology for demonstration. Fig. \ref{fig_2} (d) shows the mean-square deviation (MSD) convergence behavior under different values of $p$. When $p=0.95$ and $p=0.2$, the convergence curves are nearly identical, indicating that the method does not require almost-always-on cross-tier connections to achieve rapid consensus. Even when the activation probability is reduced to $p=0.01$, the MSD continues to decrease steadily, albeit at a slower rate. For the extremely sparse case of $p=0.001$, the algorithm remains stable; however, the lack of frequent cross-tier exchanges significantly slows global error reduction and leaves a visible residual gap within the simulated time horizon.These results demonstrate that the proposed scheme is highly robust to unreliable cross-tier communication. This corroborates Remarks 2 and 3, which establish that the network's vulnerability to cross-tier link failures does not compromise the guarantee of convergence; it solely impacts the rate of convergence. Occasional successful ground-to-LEO contacts are sufficient to inject global information into the LEO-assisted backbone, while intra-tier mixing sustains local information diffusion between these cross-tier activations. Therefore, the \textit{Sparse global + LEO} architecture can tolerate severe cross-link intermittency, with performance degrading gracefully as $p$ decreases.

\section{Discussion}
The results presented in this study demonstrate that the reliance on ultra-dense, geographically constrained terrestrial infrastructure is no longer a strict prerequisite for generating high-precision GNSS products. By adopting the proposed decentralized processing architecture, the network effectively mitigates the inherent bottlenecks of ground-to-space communication—namely, link intermittency, asymmetry, and limited bandwidth—through strategic, opportunistic cross-tier exchanges. The observed stability in LEO orbit and clock estimation, alongside resilient PPP and PPP-RTK performance under lossy conditions, underscores the viability of treating spaceborne and terrestrial stations as collaborative, dynamic subnetworks. Ultimately, this paradigm shift not only substantially reduces communication overhead compared to traditional centralized baselines but also paves the way for globally consistent, scalable, and autonomous GNSS augmentation services that are fundamentally robust to physical network constraints.










\section*{acknowledgements}
This work was supported by the King Abdullah University of Science and Technology (KAUST) Office of Sponsored Research (OSR) under Award No. RFS-CRG12-2024-6478.

\bibliographystyle{apalike}
\bibliography{references}

\end{document}